\documentstyle[12pt]{article}
\begin{document}
\noindent
\begin{center}
{\Large {\bf Conformal Invariance,\\ Anomalous Gravitational
Coupling and Derivation of a Particle Concept\\}} \vspace{2cm}
${\bf Hadi~Salehi}$
\footnote{e-mail:~h-salehi@cc.sbu.ac.ir.} \\
\vspace{0.5cm} {\small {Institute for Studies in Nonlinear
Analysis, School of Mathematical Sciences, \\Shahid Beheshti
University, P.O.Box 19395-4716, Evin, Tehran 19834, Iran.}}\\
and\\ {\small {Department of Physics, Shahid Beheshti University,
Evin,
Tehran 19839,  Iran.}}\\
\vspace{1cm} ${\bf Yousef~
Bisabr}$\footnote{e-mail:~y-bisabr@srttu.edu.}\\
\vspace{.5cm} {\small{Department of Physics, Shahid Rajaee
University,
Lavizan, Tehran 16788, Iran.}}\\
\end{center}
\vspace{1cm}
\begin{abstract}

We study a gravitational model whose vacuum sector is invariant
under conformal transformations.  In this model we investigate
the anomalous gravitational coupling of the large-scale matter.
In this kind of coupling the large-sale matter is taken to couple
to a metric which is different but conformally related to the
metric appearing explicitly in the vacuum sector. The effect of
the conformal symmetry breaking of the large-scale matter would
lead in general to a variable strength of the anomalous
gravitational coupling. This feature is used to derive a
relativistic particle concept which shares the essential
dynamical characteristics of the particle concept used in the
causal interpretation of quantum mechanics with respect to the
form of the Hamilton-Jacobi equation. The basic aspect of this
result is that it relates the variable strength of the anomalous
gravitational coupling of the large-scale matter to the
appearance of a term similar to a quantum potential term. Some of
the general characteristics of the corresponding pilot wave are
discussed.
\end{abstract}
\vspace{3cm}
\newpage
The principle of conformal invariance requires that all the
fundamental equations of physics should be invariant under local
(spacetime dependent) changes of units of length and time
\cite{bm}.  This requirement poses a fundamental symmetry in
physical theories which is of ever-increasing interest.  The
establishment of such a symmetry in the vacuum sector of a
gravitational model leads one to face with a problem concerning
the incorporation of matter to such models.  In fact, once the
vacuum sector is taken to be conformally invariant, all the
conformal frames of the metric should be considered as
dynamically equivalent. By implication there is no $a~ priori$
evidence determining to which of these frames the matter
couples.  From a classical point of view one may appeal to the
weak equivalence principle which forces us to consider the
coupling of matter to the metric which enters explicitly the
action describing the vacuum sector. This is what is generally
known as the normal coupling. At the quantum level, however,
there is not a direct evidence demonstrating the validity of the
weak equivalence principle. Thus questions such as what is the
nature of gravitational coupling of matter at this level and what
are the physical consequences of this coupling are of obvious
theoretical significance.

In this paper we deal with questions of this type
in the framework of a conformal invariant gravitational model in
which the gravitational interaction of matter can be described
both by the normal coupling and the anomalous coupling.  The anomalous
coupling means that the metric tensor entering the gravitational
part and that entering the matter part of the action correspond
to different conformal frames. We shall establish such a
distinction by introducing a nontrivial conformal factor which in
our model is considered as a dynamical field.
We show that the emergence of this field can be used to model a
relativistic particle concept which shares the essential
characteristics of the particle concept in the casual
interpretation of quantum mechanics \cite{bh} with respect to the form
of the Hamilton Jacobi equation.
This deduction of
the particle concept is carried out on the basis of a conformal
symmetry breaking.

It should be noted that the relation between the conformal
invariant gravitational models and causal interpretation of
quantum mechanics has been studied in many works, see, e.g
\cite{wp1} \cite{wp2} \cite{ssd}. The present work deals with the
exploration of the role played by the anomalous gravitational
coupling in this context. Throughout the following we shall use
units in which $\hbar=c=1$ and the signature is (-+++).

Consider the vacuum sector of a scalar tensor theory defined
by the conformal invariant gravitational action functional
\begin{equation}
S=\frac{1}{2} \int d^{4}x \sqrt{-g}~ (g^{\mu\nu} \nabla_{\mu}\phi
\nabla_{\nu}\phi+\frac{1}{6} R \phi^{2})~,
\label{a1}\end{equation} where $\phi$ is a scalar field, $R$ is
the curvature scalar associated with the metric tensor $g_{\mu
\nu}$ and $\nabla_{\mu}$ denotes a covariant differentiation.

The conformal invariance of the action (\ref{a1}) means that it is
invariant under conformal transformations
\begin{equation}
g_{\mu\nu} \rightarrow \Omega^2  g_{\mu\nu}~,
\label{2}\end{equation}
\begin{equation}
\phi(x) \rightarrow  \Omega^{-1}  \phi(x)~,
\label{3}\end{equation} where $\Omega$ is a smooth dimensionless
spacetime function. The implication is that in the vacuum sector
of a gravitational model defined by the action (\ref{a1}) the
conformal frame of the metric remains dynamically unspecified.
This point is very important concerning the gravitational
coupling of a matter system. In fact if one wants to add a matter
action to the gravitational action (\ref{a1}) one has to decide
which conformal frame of the metric is chosen for the description
of the matter action. The dynamical characteristics of a matter
system may depend significantly on the choice of this conformal
frame.

To incorporate a matter system one may assume that the conformal
frame of the metric used in the matter action is the same as that
of the gravitational action (\ref{a1}) . This corresponds to
the normal gravitational coupling and reflects one of the main
characteristics in the standard description of matter systems.
For our purpose we need to consider besides the normal coupling
the anomalous coupling in which case one distinguishes between
the conformal frame of the metric used in the gravitational
action (\ref{a1}) and that used in the matter action. In explicit
terms we intend to investigate a biscalar-tensor theory defined
by the action functional
\begin{equation}
S^{'}=\frac{1}{2} \int d^{4}x \sqrt{-g}~ \{g^{\mu\nu}
\nabla_{\mu}\phi \nabla_{\nu}\phi+(g^{\mu\nu}\nabla_{\mu}\sigma
\nabla_{\nu}\sigma +\frac{1}{6} R)\phi^{2} +\frac{1}{2}\lambda
\sigma \phi^4 \}+S_{m}[\psi,\bar{g}_{\mu \nu}]~,
\label{4}\end{equation} in which $\bar{g}_{\mu
\nu}=e^{\sigma}g_{\mu\nu}$ and $S_{m}[\psi,\bar{g}_{\mu \nu}]$
stands for the matter action which contains some matter field
variables, collectively denoted by $\psi$, with anomalous
gravitational coupling. The new scalar field $\sigma$ describes
the distinction between the two conformal frames of the metric
describing the gravitational and the matter part of the action
(\ref{4}). The presence of a kinetic term for $\sigma$ implies
that $\sigma$ is considered as a dynamical field. The action
(\ref{4}) contains also an interaction term $\sim \lambda \sigma
\phi^4$ in which $\lambda$ is a dimensionless coupling constant.
The particular form of this interaction term and the kinetic term
of $\sigma$ is suggested by the fact that these terms respect the
conformal symmetry of the gravitational action
(\ref{a1})\footnote{Note that $\sigma$ as a dimensionless
function does not transform under a conformal transformation.}.
Thus the total action (\ref{4}) is distinguished by the fact that
a conceivable (conformal) symmetry breaking may be linked to the
physical characteristics of the matter at hand.

At this point it is necessary to clarify the notion of the
anomalous gravitational coupling introduced by the action
(\ref{4}) with respect to other existing publications devoted to
this issue. Usually one uses conformal transformations of the
metric to transfer the action of a scalar-tensor theory into
different inequivalent representations, for a review see
\cite{far}. This is done, for example, in Brans-Dicke theory to
represent the corresponding action in the so called Einstein
frame in which a constant strength of the gravitational coupling
can be established. In this approach there is no a priori
distinction between the conformal frame of the metric used in the
gravitational part and that used in the matter part of the
original action. The anomalous gravitational coupling emerges in
all these cases because the applied conformal transformation
generates an exponential factor before the metric \cite{da} in the
matter action or occasionally before the matter part of the
Lagrangian \cite{far}. The notion of the anomalous coupling
introduced by the action (\ref{4}) is however conceptually
different with respect to two points. Firstly, it allows us to
have an a priori distinction between the two conformal frames of
the metric used in the gravitational part and in the matter part
of the original action. Secondly, the action is constructed in
such a way that the gravitational part of the action exhibits the
conformal symmetry so that the physical characteristics of the
matter may be considered as responsible for a conceivable
conformal symmetry breaking. \footnote{Note that the class of
scalar tensor theories which is mostly discussed in the context
of the anomalous coupling breaks the conformal symmetry already
in the vacuum sector, i.e., at the level of the gravitational
part of the original action.}. This last point has important
implications regarding a new symmetry breaking phase if the
matter actually breaks the conformal invariance. In such cases
the anomalous gravitational coupling of matter can be related to
an anomalous phase of the symmetry breaking. This anomalous phase
is often ignored in the consideration of conformal symmetry
breaking \cite{ds} which is usually based on the normal coupling
(the normal phase). We shall deal with this issue later in
connection with the symmetry breaking effect due to the
gravitational coupling of the large-scale matter.

Varying $S^{'}$ with respect to $g^{\mu \nu}$, $\phi$ and $\sigma$
yields, respectively,
\begin{equation}
G_{\mu \nu}-\frac{3}{2}\lambda \sigma \phi^2 g_{\mu\nu}
=6\phi^{-2}(e^{\sigma}\Sigma_{\mu \nu}+\tau_{\mu\nu})+ 6t_{\mu\nu}~, \label{5}\end{equation}
\begin{equation}
\Box \phi-\frac{1}{6}R \phi-\lambda \sigma \phi^3- \phi
\nabla_{\alpha }\sigma \nabla^{\alpha  }\sigma =0~,
\label{6}\end{equation}
\begin{equation}
\nabla_{\mu}J^{\mu} =\frac{1}{2}\sqrt{-g}(\frac{1}{2}\lambda
\phi^4 +e^{\sigma}g^{\mu\nu}\Sigma_{\mu\nu})~,
\label{6a}\end{equation} where
\begin{equation}
\Sigma_{\mu \nu}=\frac{-2}{\sqrt{\bar{g}}} \frac{\delta}{\delta
\bar{g}^{\mu \nu} } S_{m}[\psi,\bar{g}_{\mu \nu}]~,
\label{6b}\end{equation}
\begin{equation}
J^{\mu}=\sqrt{-g}\phi^2 g^{\mu\nu}\nabla_{\nu}\sigma~,
\label{6aa}\end{equation} and
\begin{equation}
\tau_{\mu \nu}= -(\nabla_{\mu }
\phi \nabla_{\nu}\phi-\frac{1}{2}g_{\mu \nu} \nabla_{\alpha  }\phi
\nabla^{\alpha  }\phi)-\frac{1}{6}(g_{\mu \nu} \Box-\nabla_{\mu }
\nabla_{\nu})\phi^2~,
\end{equation}
\begin{equation}
t_{\mu \nu}=-\nabla_{\mu }\sigma \nabla_{\nu}\sigma +\frac{1}{2}
g_{\mu \nu} \nabla_{\alpha  }\sigma \nabla^{\alpha }\sigma~.
\label{7}\end{equation} Here $\Box \equiv g^{\mu \nu}
\nabla_{\mu} \nabla_{\nu}$ and $G_{\mu \nu}$ is the Einstein
tensor of the metric $g_{\mu \nu}$.  In the equation (\ref{5}) we
consider $\Sigma_{\mu\nu}$ as the matter stress tensor. Note that
this tensor appears in (\ref{5}) with a conformal weight
$e^{\sigma}$. This is a consequence of the anomalous coupling
which forces us to define $\Sigma_{\mu\nu}$ in terms of
functional differentiation with respect to
$\bar{g}_{\mu\nu}$\footnote{In the present context, there may be
an ambiguity in the definition of the conformal weight of the
matter stress tensor when matter systems anomalously couple to
gravity. This ambiguity does not affect the basic result of the
present work. The exploration of the issue is however deserved to
be investigated elsewhere.}.

Taking the four-divergence of (\ref{5}) and combining the result
with (\ref{6}) and (\ref{6a}) leads to
\begin{equation}
\nabla^{\mu }(e^{\sigma}\Sigma_{\mu \nu})
=\frac{1}{2}\nabla_{\nu}\sigma~ e^{\sigma} ~g^{\mu\nu}
\Sigma_{\mu\nu}~, \label{7a}\end{equation}
which indicates that $\Sigma_{\mu \nu}$ can
not be conserved due to the presence of $\sigma$.  We also note
that in the absence of matter the trace of (\ref{5}) would lead
directly to the equation (\ref{6}) and, therefore, the latter
would not contain any new information. This is a direct
consequence of conformal invariance of the vacuum sector of
(\ref{4}).  In the presence of matter, on the other hand, we
obtain the following condition
\begin{equation}
e^{\sigma}g^{\mu\nu}\Sigma_{\mu\nu}=0~,
\label{ab}\end{equation} by comparing the trace of (\ref{5}) with
the equation (\ref{6}). This is a dynamical consistency relation
imposed on the behavior of matter systems in this model. It means that
only the gravitational coupling of
traceless matter systems is allowed to be described by
(\ref{4}). To describe the gravitational coupling of
matter systems with nonvanishing
trace of the corresponding stress-tensor one may add a term
such as $\frac{1}{2} \int d^{4} x \sqrt{-g} \mu^2 \phi^2$ to the
total action in (\ref{4}) with $\mu$ being a constant mass parameter.
In this case the
analog of the equations (\ref{5}) and (\ref{6}) are
\begin{equation}
G_{\mu \nu}-3\mu^2 g_{\mu\nu}-\frac{3}{2}\lambda \sigma \phi^2
g_{\mu\nu} =6\phi^{-2}(e^{\sigma}\Sigma_{\mu \nu}+\tau_{\mu\nu})+ 6t_{\mu\nu}~, \label{5-a}\end{equation}
\begin{equation}
\Box \phi-\frac{1}{6}R \phi-\lambda \sigma \phi^3- \mu^2 \phi-\phi
\nabla_{\alpha }\sigma \nabla^{\alpha  }\sigma =0~,
\label{6-a}\end{equation} and the equation (\ref{6a}) remains
unchanged. The relation (\ref{ab}) takes then the form
\begin{equation}
e^{\sigma}g^{\mu\nu}\Sigma_{\mu\nu}=-\mu^2
\phi^2~. \label{ac}\end{equation} This ensures the consideration
of the gravitational coupling of
matter systems with nonvanishing trace of the corresponding
stress-tensor to be dynamically
consistent.

It should however be noted that
the appearance of a constant dimensional parameter such as $\mu$
clearly breaks the conformal symmetry. We are primarily interested
in the investigation of a cosmological symmetry breaking emerging from the
gravitational coupling of the large-scale distribution of matter
in the universe. In this case the length scale $\mu^{-1}$ should
be related to the typical size of the universe $R_0$, namely
$\mu^{-1}\sim R_0$. We first consider the normal phase of the
symmetry breaking which is taken to be based on the normal
coupling of the large-scale distribution of matter. This demands
a vanishing average value for $\sigma$ at large scales, namely
$\bar{\sigma}=0$. It this case it is possible to obtain an
estimation for the constant background average value of $\phi$
which provides the strength of the gravitational coupling. In
fact, the trace of $\Sigma_{\mu\nu}$ can be measured in terms of
the average density of the large-scale distribution of matter,
i.e., $\Sigma_{\mu}^{\mu} \sim -M/R_{0}^3$ which $M$ denotes the
mass of the universe.  If one uses the empirical fact that the
radius of the universe coincides with its schwarzschild radius
$2GM$, one then gets an estimation of the constant background
value of $\phi$
\begin{equation}
\phi^{-2} \sim  G \sim m_{p}^{-2}~, \label{G}\end{equation} where
$G$ and $m_{p}$ are the gravitational constant and the Planck
mass, respectively. In this case, the gravitational equation
(\ref{5-a}) reduces to the Einstein field equations with a small
cosmological constant.  It is important to note that the use of
$\lambda \sigma \phi^4$, instead of the usual $\lambda \phi^4$
term, allows
to avoid the cosmological constant problem in a cosmological context.\\
Thus it is possible to use a constant configuration of $\phi$ to
measure a constant strength of gravitational coupling of the
large scale distribution of matter in the normal coupling. It
should be emphasized that this feature seems to be characteristic
only to the normal phase of the symmetry breaking which is based on the normal
coupling.
In the case of anomalous coupling
the conformal symmetry breaking can not be used to achieve a
precise estimation of the energy density of
$\Sigma_{\mu\nu}$. This means that we may not
obtain any knowledge of the configuration of the scalar
field $\phi$ as a measure for the strength of the anomalous gravitational
coupling.

We proceed now to the consideration of the anomalous phase of
the symmetry breaking which is taken to be based on the anomalous coupling of
the large-scale distribution of matter. The corresponding
symmetry breaking is characterized by the relation (\ref{ac}) in
which the length scale $\mu^{-1}$ again measures the typical size
of the universe. But (\ref{ac}) implies that the local variations
of $\sigma$ lead to a variable configuration of $\phi$. As a
consequence, the strength of the anomalous gravitational coupling
of the large scale distribution of matter should not be
considered as constant. This feature
can be used to derive a particle concept in the anomalous phase
of the symmetry breaking. From (\ref{6-a})
we obtain
\begin{equation}
g^{\mu\nu}\nabla_{\mu}\sigma \nabla_{\nu}\sigma =-m^2-\frac{1}{6}R
+\frac{\Box \phi}{\phi}~, \label{14}\end{equation} which may be
interpreted as a generalized Hamilton-Jacobi equation of a
relativistic particle with a variable mass $m=(\lambda \sigma
\phi^2+ \mu^2)^{1/2}$. The major characteristic of (\ref{14})
lies in the term $\frac{\Box \phi}{\phi}$.  It is similar to the
term related to the so-called quantum potential in the context of
causal interpretation of relativistic quantum mechanics
\cite{bh}.  In present context, this term emerges because the
anomalous gravitational coupling of large-scale matter systems
does not provide us a constant strength for the gravitational
coupling.

In order for the particle interpretation of (\ref{14}) to be
valid two general characteristics should be present. Firstly the
particle trajectories must be timelike. Secondly the variable
mass scale $m$ must remain constant along the particle
trajectories. To establish the first characteristic one may
impose the condition $\nabla_{\alpha }\sigma \nabla^{\alpha
}\sigma < 0$ to ensure that the particles move along timelike
trajectories. Looking at (\ref{7}) we see that this acts as an
energy condition on the stress tensor $t_{\mu\nu}$. It selects,
namely, those configurations of $\sigma$ for which the trace of
$t_{\mu\nu}$ is negative. For the stress tensors of matter
systems this assumption is dynamically consistent, as one may
infer from (\ref{ac}). Thus the restriction to timelike
trajectories means that the stress tensor $t_{\mu\nu}$ which
appears on the right hand side of (\ref{5-a}) follows the general
behavior of dynamically allowed matter systems with respect to
the sign of the trace. The establishment of the second
characteristic needs careful considerations. The constant mass
scale of a particle seems to be a feature which is valid if a
particle could ideally be considered as a quasi-isolated object.
To perform the corresponding limiting procedure we propose to
consider the scaling limit $\mu\rightarrow 0$, i.e., the
idealized limit of infinite size of the universe. In this limit
the conformal invariance is restored. Taking this limit as
reasonable we may then select a conformal frame (a particle
frame) by the condition
\begin{equation}
\nabla_{\alpha}\sigma \nabla^{\alpha }m=0~,
\label{9}\end{equation} which ensures that the mass scale $m$
remains constant along the particle trajectories\footnote{The
masses of the particles contribute to vacuum energy density and
appear as a large cosmological constant in the gravitational
equation (\ref{5-a}). Thus the question arises whether this would
lead to a cosmological constant problem \cite{w}. This issue is
difficult to be resolved, but we make here only a modest remark.
In the present case this problem arises only in the anomalous
phase of symmetry breaking. In fact the mass of the particles is
generated by a nonvanishing value of $\sigma$ which in our
presentation is characteristic to the anomalous coupling of the
large-scale matter. However the considerations in cosmology are
offer based on the normal coupling of the large-scale matter. The
vanishing value of $\sigma$ ensures in this case that no
cosmological constant problem arises. One can probably say that
the cosmological constant problem arises if one makes no
distinction between the nature of gravitational coupling of
matter in cosmology and particle physics \cite{bs}. The issue is
however too difficult to be resolved within the scope of this
paper.}.

Finally we turn to the general particle concept
described by (\ref{14}) and remark that the motion of the
particles can be regarded as influenced by a guidance wave, or a
pilot wave, defined by
\begin{equation}
\psi=\phi e^{i\sigma }~. \label{15}\end{equation} Using
(\ref{6aa}) and (\ref{14}), one can easily show that $\psi$
satisfies the wave equation
\begin{equation}
(\Box-m^2-\frac{1}{6}R)\psi=\frac{i\phi^{-2}}{\sqrt{-g}}~\nabla_{\mu}J^{\mu}~~\psi~.
\end{equation}
As a consequence of (\ref{6a}) and in the scaling limit $\mu
\rightarrow 0$, this reduces to
\begin{equation}
(\Box-m^2-\frac{1}{6}R)\psi=i\frac{m^2}{4\sigma}\psi~.
\label{cc1}\end{equation}
The right hand side of this equation
indicates that there is an imaginary contribution to the squared
mass of the particles. However one expects that such a
contribution to be vanishingly small since in order to achieve a
particle interpretation, there should exist a precise distinction
between the two conformal frames of the metric
describing the gravitational and the
matter part of the action (\ref{4}). Thus one may require that
$\sigma$ takes sufficiently large values. In this case the right hand side
of the equation (\ref{cc1}) drops out and this equation takes the
form of the usual field equation in a gravitational background.

We summarize our observations as follows: A breakdown of conformal
invariance of the gravitational model (\ref{4}) in the anomalous
phase provides a theoretical framework by which one may derive
a particle concept.  This derivation entails the existence of a
dynamical field which describes the anomalous coupling of the
large-scale matter, namely the scalar function $\sigma$. This
scalar function makes it possible to have a distinction between
the conformal frame of the metric used in the description of the
large-scale matter system and that used in the description of the
geometry of spacetime.  In the absence  of this distinction (the
normal phase of symmetry breaking) the theory describes just the
usual gravitational coupling of large-scale distribution of
matter.  In the presence of this distinction (the anomalous phase
of symmetry breaking), on the other hand, the theory can be used
to model local particle systems.  In this case it is essentially
the local variation of the scalar function $\sigma$ that leads
one to a particle interpretation in which $\sigma$ plays the role
of the Hamilton-Jacobi function. This demonstrates the nontrivial
role of the absolute value of the phase of a wave function in the
gravitational context. The local variation of $\sigma$ has
another important implication in that it predicts a variable
strength of the gravitational coupling of the large-scale matter.
This leads to the appearance of a corresponding term in the
Hamilton-Jacobi equation which is similar to the quantum
potential term used in the causal interpretation of quantum
mechanics. The whole derivation of the particle concept
suggests that the anomalous phase of the symmetry breaking
due to the anomalous gravitational coupling of the large-scale matter
may have a vital role to play in the geometric
interpretation of causal framework of quantum mechanics.


\end{document}